\documentstyle[sprocl]{article}

\input{psfig}

\def\Journal#1#2#3#4{{#1} {\bf #2}, #3 (#4)}

\def\NPB{{\em Nucl.\ Phys.} B}
\def\PLB{{\em Phys.\ Lett.}  B}
\def\PRL{\em Phys.\ Rev.\ Lett.}
\def\PRC{{\em Phys.\ Rev.} C}
\def\PRD{{\em Phys.\ Rev.} D}

\def\be{\begin{equation}}
\def\ee{\end{equation}}
\def\bea{\begin{eqnarray}}
\def\eea{\end{eqnarray}}
\begin{document}

\mbox{ } \vspace{-0.5in}

\vbox{\hfill\hbox{JLAB-THY 98--38}} \vspace{0.3in}

\title{NN INTERACTIONS IN QCD: OLD AND NEW TECHNIQUES\footnote{Invited
talk presented at {\it Mesons and Light Nuclei}, Aug.\ 31--Sept.\ 4,
1998, \mbox{Pru\hspace{-1.2ex}\raisebox{1.2ex}{\tiny $\circ$}honice}
(Prague), Czech Republic; to appear in Proceedings.}}

\author{Richard F. Lebed}

\address{Jefferson Lab, 12000 Jefferson Avenue \\ 
Newport News, VA 23606, USA\\E-mail: lebed@jlab.org} 


\maketitle\abstracts{Since QCD is believed to be the underlying theory
of the strong interaction, it is appropriate to study techniques that
take into account more features of its rich and complex structure.  We
begin by discussing aspects of physics that are ill-reproduced by the
usual one- or two-meson exchange approaches and identify the source of
the deficiencies of these models.  We then reveal promising methods
for curing some of these ills, such as new quark potential models,
baryon chiral perturbation theory, soluble strongly-interacting field
theories, and large $N_c$ QCD.}

\section{Introduction}

	As a particle theorist, I was invited to speak at this
conference about what can be said regarding the nucleon-nucleon
interaction in the context of quantum chromodynamics.  Hadronic
physics is probably the most difficult problem in the entire panoply
of particle theory, and the primary quest of its practitioners (myself
included) is to uncover some particularly simple physical picture that
not only respects the physics of observed hadrons, but arises as a
natural consequence of the substructure of quarks and gluons.  To this
day we remain hindered in our ability to provide an eloquent and
definitive solution to the problem.

	However, even on that day of QCD's eventual solution, the
meson exchange picture of nuclear physics will remain the natural
picture for the NN interaction in almost all situations, in precisely
the same way that NASA quite successfully employs Newtonian
gravitation for calculating trajectories of spacecraft, the existence
of general relativity as a more ``fundamental'' theory of gravitation
notwithstanding.  Quarks and gluons are certainly present in all NN
interactions, but it is not always necessary to take them explicitly
into account.

	Nevertheless, we have learned quite a bit about QCD in its
first quarter century.  We know the Lagrangian and its symmetries,
some properties of the quarks themselves, and a bit about the nature
of color confinement.  Moreover, one can study versions of QCD-like
theories that have been simplified so that some of the difficult
physics becomes tractable.  These sorts of advances can be applied to
the NN problem, to constrain types of possible dynamics or to reduce
allowed parameter spaces.  The significance of such physics becomes
most apparent when one considers circumstances in which the standard
meson exchange picture begins to falter, so we begin with a discussion
of the successes and problems of this picture.  We then consider
exactly what it means to obtain QCD improvements, and exhibit a number
of techniques designed to accomplish this goal.

\section{The Age of the Meson Exchange Picture}

	The NN interaction is perhaps the most studied of all problems
in nuclear physics, and the decades of careful scrutiny and hard work
are now coming to fruition in the convincing numerical success of
various famous potential models.  With only a score or two of
parameters, the models of Paris,~\cite{paris} Bonn,~\cite{bonn}
Nijmegen,~\cite{nij} and Argonne~\cite{arg} have begun to achieve a
global fit to the numerous experimentally observed partial waves and
hundreds of extracted data points approaching the all-important
statistical criterion of $\chi^2/{\rm d.o.f.} \approx 1$.  The
fundamental physics incorporated in these models is obtained from the
current understanding of the dynamics of meson exchange: The unknown
fit parameters appear in the context of Yukawa potentials, form
factors, and so forth.

	Are we then to conclude that the NN interaction is essentially
a solved problem, with the few remaining discrepancies requiring only
minor adjustments of functional forms or numerical values of the
parameters?  From a strictly reductionist point of view, as the fits
of models to data become ever more precise, questions about the origin
of these numerical values become more pressing.  Moreover, one has in
QCD the apparent underlying fundamental theory of strong interactions.
Even if one looks askance at future prospects of describing a nuclear
problem such as the NN interaction with its complicated phenomenology
in the language of quarks and gluons, one cannot believe that all of
the masses and couplings of the common mesons are independent
quantities.  The proper question is, how do nuclear phenomena arise as
a limit of QCD?

	In the traditional picture of the NN potential, at large
internuclear separations (\raisebox{-0.5ex}{$\stackrel{>}{\sim}$} 3
fm) one observes exponential saturation of nuclear forces with a
residual attraction, which is explained by one-pion exchange. At
intermediate ranges, between roughly 0.6 and 3 fm, there is an
attractive region typically ascribed to exchange of a scalar $\sigma$
meson or scalar combination of two $\pi$'s, while at smaller distances
one finds an effective short-range repulsion, which is identified with
$\omega$ exchange.  Including the $\rho$, whose primary role is to
cancel most of the tensor interaction of the $\pi$ at short distance,
completes this picture.  One is immediately faced with questions about
the identity of these mesons, $\pi$, $\sigma$, $\rho$, $\omega$.  Are
they the fundamental mesons appearing in the Particle Data Book, or do
they merely serve as placeholders to parametrize more complicated
physics?  For example, the troublesome ``$\sigma$'' might actually
indicate the exchange of scalar current between nucleons arising from
many different sources at different momentum scales, including but not
limited to $\pi \pi$ pairs.

\begin{figure}[t]
%\rule{5cm}{0.2mm}\hfill\rule{5cm}{0.2mm}
%\vskip 2.5cm
%\rule{5cm}{0.2mm}\hfill\rule{5cm}{0.2mm}
%\hspace{0.875in}
\hfil \psfig{figure=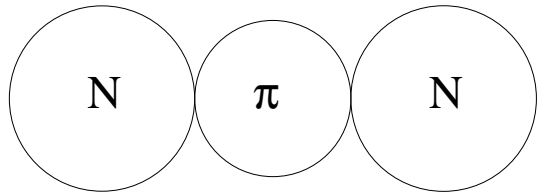,height=1.0in} \hfil
\caption{A parable for the short-distance failure of one-meson
exchange in the NN interaction.\label{fig:mech}}
\end{figure}

	The fundamental problem with one-meson exchanges as a
universal explanation for NN phenomena can be described in terms of
the following simple ``parable'' (Fig.~\ref{fig:mech}): Think of
hadrons as classical hard spheres.  The charge radius of the proton is
$\sqrt{ \langle r^2 \rangle} \approx 0.6$ fm, while the typical size
of light mesons is set by the QCD scale $\Lambda_{\rm QCD}$, $r
\approx 1/\Lambda_{\rm QCD} \approx 1/(300 {\, \rm MeV}) = O$(0.5 fm).
Then a meson cannot mediate an NN interaction at distances below (0.6
fm + 2 $\cdot$ 0.5 fm + 0.6 fm) = 2.2 fm: It simply won't fit between
the nuclei!
Of course, real hadrons are quantum mechanical, but the point
of the parable remains: One-meson exchange only makes sense inasmuch
as the meson (and, to a smaller extent, the nucleon) can be treated as
a point particle.  At distances smaller than a couple of fermi, one
probes the inner structure of the hadrons, which is to say their short
distance or high momentum components.  In meson models this is
typically taken into account with form factors.  But which type of
form factor is correct?  Obviously, this is a situation in which more
input from QCD would provide valuable clarification.

\begin{figure}[t]
%\rule{5cm}{0.2mm}\hfill\rule{5cm}{0.2mm}
%\vskip 2.5cm
%\rule{5cm}{0.2mm}\hfill\rule{5cm}{0.2mm}
%\hspace{0.5in}
\hfil \psfig{figure=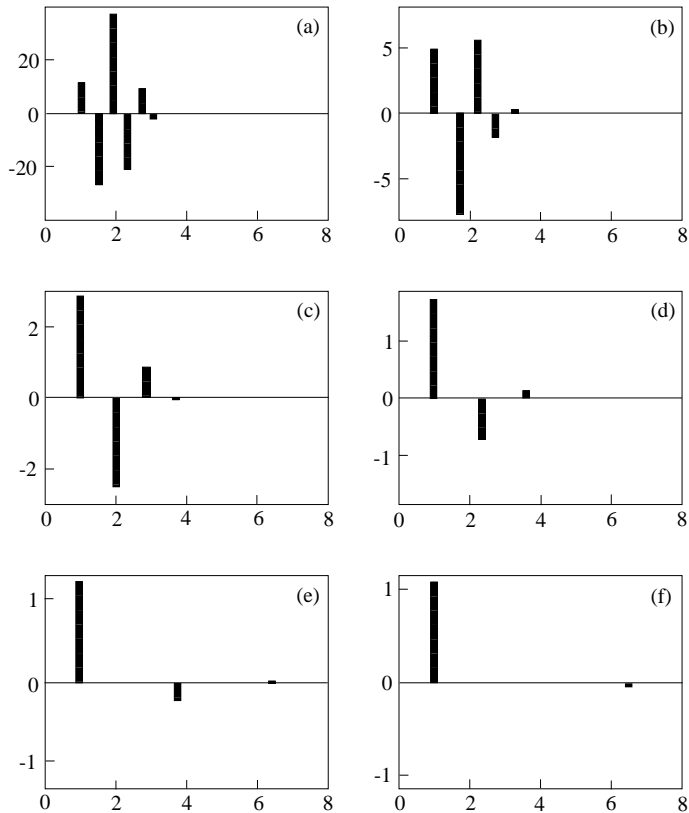,height=4.25in} \hfil
%\vskip 4.25in
\caption{Size of form factor pole residue contributions $\Lambda_n$
vs.\ $\mu_n^2/\mu_0^2$, shown for quark masses decreasing in the order
$(a)$--$(f)$ (after Ref.~$^6$).\label{fig:JMfig}}
\end{figure}

	Perhaps you are unconvinced by this parable, and would prefer
a more direct demonstration of some situation where one-meson exchange
fails.  To be suitably rigorous, one needs a theory in which quark and
hadron degrees of freedom can be handled equally well.  There in fact
exists a completely soluble, strongly interacting theory, which is
called the 't~Hooft model;~\cite{tHmodel} by definition, it is QCD in
one space and one time dimension with a large number of color charges
$N_c$.  ``Completely soluble'' here means that one can obtain meson
masses, wavefunctions, and transition amplitudes precisely in terms of
quark masses.  Within this framework, consider a typical hadronic
quantity, the meson electromagnetic form factor:~\cite{JM}
\begin{equation}
F(q^2) = \sum_{n=0}^\infty \frac{\Lambda_n \mu_n^2}{q^2 - \mu_n^2 + i
\epsilon} ,
\end{equation}
where $\mu_n$ and $\Lambda_n$ are the mass and pole residue of the
$n$th meson.  That the form factor may be written as a sum of pole
terms is a consequence of large $N_c$.  Single-meson exchange models
tell us to expect $\Lambda_0 \gg \Lambda_1, \Lambda_2, \ldots$, {\it
i.e.,} that the contribution of the lightest meson is the most
important.  Carrying out the calculation in the 't~Hooft model, one
finds that this is true only for light quarks (Fig.~\ref{fig:JMfig}).
As the quark mass increases, one sees not only that a larger number of
poles become significant, but that they alternate in the signs of
their residues, meaning that a one-meson exchange picture becomes
progressively more inadequate.  Indeed, as the quark mass becomes very
large, one can show that the residues arrange themselves to give the
same predictions as the nonrelativistic quark model.

\section{Including More QCD}

	Since we have argued that there are circumstances in which
input from QCD is essential for physical understanding, one must
define exactly what is meant by QCD improvement.  Let us adopt the
broadest definition possible, in order to represent a complete
spectrum of the many features of the strong interaction.  Then QCD
improvements fall into two basic categories.  First, since QCD is a
quantum field theory, it must satisfy the properties of causality,
relativistic covariance, crossing symmetry, unitarity, and
unrestricted production of virtual particles.  Second, a number of
field-theoretic properties are special to QCD, namely, the presence of
quarks and gluons (of course), as well as gauge invariance and color
conservation, color confinement, and the discrete symmetries of
parity, charge conjugation, and time reversal invariance.  In
addition, QCD provides for many kinds of hadrons with numerous types
of nonlinear interactions, and --- very importantly --- approximate
chiral symmetry.

\subsection{Inverse Scattering}

	First consider an improvement that takes into account only
field theoretic QCD properties.  The premise of inverse scattering is
that one uses data directly from the $S$-matrix and phase shifts as
functions of momentum transfer $k$ via
\begin{equation} \label{smat}
S_\ell (k) = \exp \left( 2i \delta_\ell (k) \right) ,
\end{equation}
which, by construction, automatically satisfies field theoretic
aspects of QCD, since the $S$-matrix obeys unitarity, crossing
symmetry, and so forth.  One then inverts (\ref{smat}) using standard
mathematical techniques such as Gel'fand-Levitan or Marchenko
inversions,~\cite{Newt} to obtain an equivalent {\em local\/} potential
$V(r)$ that, by construction, agrees with the data.  If one then
compares to usual potential models, the agreement is quite good for
most partial waves,~\cite{KABCG} since these potentials were designed
expressly to fit the phase shifts.  In particular, $V(r)$ obtained in
this method exhibits a repulsive core.  Moreover, the inversion may be
continued off-shell to produce interesting results relevant to
processes like nucleon bremsstrahlung.~\cite{JSG} However, a local
potential $V(r)$ depends on only one coordinate $r$, which is the
separation of the two nucleon centers.  This is a natural picture when
the nucleons may be considered (nonrelativistic) point particles, but
may be inadequate when nucleonic substructure is taken into account,
in which case there is more than just one relevant separation
coordinate.

\subsection{Local and Nonlocal Potentials}

	The nucleonic substructure of quarks and gluons can create
nonlocality in the NN potential, which may be expressed as an energy
dependence $V(E,r)$.  It is true that relativistic effects also
produce an energy-dependent potential, but one can study the effects
of substructure separately by considering a nonrelativistic toy model.
An interesting example of this approach appears in Ref.~\cite{SY},
where the $p$-$\Sigma^+$ potential, known to be have a highly nonlocal
potential in the quark model, is considered.  The nonlocality is
introduced through a potential term
\begin{equation}
\Delta V ({\bf r}_N , {\bf r}_\Sigma ) \sim \exp \left( - \frac{\left(
{\bf r}_N + {\bf r}_\Sigma \right)^2}{4a_c^2} \right) \exp \left( -
\frac{\left( {\bf r}_N - {\bf r}_\Sigma \right)^2}{4a_d^2} \right) ,
\end{equation}
which depends on not only the separation $\left( {\bf r}_N - {\bf
r}_\Sigma \right)$ but also the average position $\left( {\bf r}_N +
{\bf r}_\Sigma \right) \! /2$.  The phase shifts obtained from this
nonlocal potential can then be used to generate an equivalent local
potential.  The result of this calculation shows that the height of
the repulsive core in the nonlocal potential is greatly reduced
compared to that from the local potential (see esp.\ their Fig.~3).
Such a conclusion suggests that the repulsive core of potential models
is actually due to nucleon substructure; it would certainly agree with
our earlier comments that single-meson exchange at short distance
should not be a good description of the NN interaction.

\subsection{Quark Models}

\begin{figure}[t]
%\rule{5cm}{0.2mm}\hfill\rule{5cm}{0.2mm}
%\vskip 2.5cm
%\rule{5cm}{0.2mm}\hfill\rule{5cm}{0.2mm}
%\hspace{0.5in}
\hfil \psfig{figure=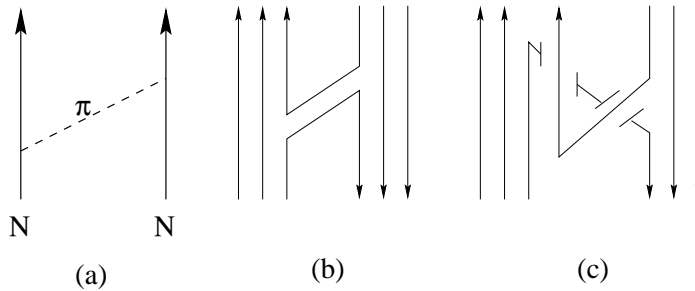,height=1.5in} \hfil
\caption{NN interactions in ($a$) meson and ($b$),($c$) quark
exchange pictures.  In ($b$) the intermediate quark exchange is easily
described as a single-meson exchange, while in ($c$) it is
not.\label{fig:color}}
\end{figure}

	One form of quark model phenomenology uses nothing more than
valence quarks either interacting in some phenomenological potential,
or with some chosen wavefunctions within the nucleon.  Such quark
model studies of features of the NN interaction have a very old
history, dating back to the dawn of the quark model itself in the mid
1960s.  The unique feature of quarks, however, is that they possess
the color degree of freedom.  Once one determines that color exists,
situations become inevitable in which the color degree of freedom must
be considered explicitly.  As an example, consider a second parable in
the form of Fig.~\ref{fig:color}.  Starting with one-meson exchanges
in the NN interaction (Fig.~\ref{fig:color}$a$), once one decides that
$N$ is a 3-quark state while mesons are $\bar q q$ states, the
interpretation of the meson exchange in terms of colored quark lines
becomes clear (Fig.~\ref{fig:color}$b$).  However, just as likely are
diagrams in which the quark lines are tangled
(Fig.~\ref{fig:color}$c$).  In such a case, the intermediate state is
clearly not a single meson, nor is it even a color singlet.  It is
still possible to describe it in the meson language, but to do so
requires a large number of carefully correlated meson exchanges; this
is the same phenomenon that we saw in the 't~Hooft model
(Fig.~\ref{fig:JMfig}) for large quark masses.  The promotion of this
argument from parable to rigor involves the inclusion as well of all
possible gluon exchanges, but it seems reasonable that such a
modification cannot completely screen all color from our notice.

	To date, however, the best quark model studies still only
include the gluon degrees of freedom through field-theoretic
reductions of one-gluon exchanges.  Nevertheless, this is enough to
capture quite a bit of physics, such as spin-orbit couplings and
hyperfine terms.  Consider, for example, the results of one particular
study,~\cite{IM} in which six-quark states are placed in an interaction
derived from single-gluon exchange plus an explicit confining
potential.  It is then found that the repulsive core originates as
anti-binding from the spin-spin coupling of the hyperfine interaction,
while the intermediate attraction is a result of the excitation of
color nonsinglet $P$-wave clusters of quarks.  Clearly, these are
phenomena that have no simple interpretation in terms of one-meson
exchange.

	Another interesting idea for the suppression of the repulsive
core arises in the context of Moscow potentials.~\cite{Mosc} One
begins with the physical observation that it is rare to observe
baryons with very small separation.  The standard explanation, of
course, is that one is seeing a potential with a repulsive core.
However, many researchers suggest that the same physics may be
obtained through an NN wavefunction with a node at small separation,
effectively suppressing such observations.  In the case of the Moscow
potential, the same sort of wavefunction suppression is achieved
through a ``two-phase'' model: Starting with six quarks, at large
separation there is a high probability for segregation into two
three-quark nucleon clusters, with combined wavefunction $\Phi$.  At
small separation there is a high probability to form a ``bag-like''
six-quark state $\Psi$.  The wavefunctions $\Phi$ and $\Psi$ are then
taken to be orthogonal.  Thus, it is not terribly difficult to push
six quarks into a very small volume, but then the dominant part of the
wavefunction no longer resembles two distinct nucleons.  Such models
allow for good fits to many of the phase shifts, and the residual
meson interaction potentials (to account for long-distance physics not
incorporated into the quark interactions) may then be taken as local.
Moreover, the $\omega$NN coupling falls to values consistent with that
predicted from SU(3) symmetry, since the $\omega$ is no longer
required to serve the special role of providing the repulsive core
interaction.

	Quark models typically explain the short- and
intermediate-distance features of the NN interaction; both nuclear and
particle physicists can agree that the long-distance tail is due to
single-pion exchange.  Nevertheless, it is fruitful to compare the
tortuous discovery of this fact with its current explanation in many
textbooks.  Historically, Yukawa proposed in 1935 the exchange of
mesons to explain nuclear binding; the $\pi$ was the first true meson
discovered (in 1947), and was subsequently found to explain the
long-distance behavior of NN interactions well.  Since the $\pi$ is
still the lightest observed meson, by the Heisenberg principle it has
the longest range.  This bottom-up process of discovery is to be
contrasted with our current top-down understanding of the same
phenomenon: The QCD Lagrangian possesses chiral symmetry, which is
spontaneously broken to an approximate flavor symmetry.  The breaking
produces a multiplet of pseudoscalar Nambu-Goldstone bosons, of which
the $\pi$ is the lightest, since it contains no heavy strange quarks.
Therefore, again by the Heisenberg principle, it should have the
largest range of any strongly interacting particle.

\subsection{Effective Theories}

	The discovery of the approximate chiral symmetry of strong
interactions has been one of the primary achievements of the extensive
efforts placed in understanding the NN interaction over the years.  It
is exploited to great effect in chiral Lagrangians and chiral
perturbation theory ($\chi$PT), and yet such theories are only one
specific type of what are now collectively called {\em effective
theories}.  Let us explain how such theories are constructed in
general, with reference to the familiar $\chi$PT case.

\begin{enumerate}

\item {\em Choose a set of fields as dynamical degrees of freedom.}
In $\chi$PT these are pions, as well as $K$'s and $\eta$'s in the
3-flavor case, and nucleons can also be incorporated into this scheme.

\item {\em Identify the symmetries obeyed by interactions of these
fields.}  In $\chi$PT these are Lorentz covariance, (approximate)
chiral symmetry, and the discrete symmetries P, C, and T.

\item {\em Express fields in forms that transform appropriately under
the given symmetries.}  For example, one convenient representation
containing the pion field \mbox{\boldmath{$\pi$}} in $\chi$PT is
$\Sigma \equiv \exp ( 2i \mbox{\boldmath{$\pi \! \cdot \! \tau$}}
/f_\pi )$, for then under SU(2)$_L \times$ SU(2)$_R$ chiral rotations
$L$ and $R$ one has $\Sigma \to L \Sigma R^\dagger$.  Here
\mbox{\boldmath{$\tau$}} are the isospin generators and $f_\pi$ is the
pion decay constant.

\item {\em Construct the Lagrangian that explicitly obeys all
symmetries.}  In general, this procedure produces an infinite list of
terms, which gives the initial naive impression that the theory has no
predictive power at all.  In the case of $\chi$PT, however, more
complicated terms with more fields or derivatives enter, by virtue of
simple dimensional analysis, with more inverse powers of some
characteristic mass scale $\Lambda$.  In practice, $\Lambda$ for
$\chi$PT is typically taken at the scale of $m_\rho$ or 1 GeV, where
describing physics solely in terms of pion interactions is no longer
adequate.  Therefore, all but a small finite number of the possible
infinite set are insignificant for a given physical process.  In the
general case, an effective theory is useful if the more complicated
terms are suppressed numerically in physical quantities, which means
that the characteristic momenta of the process must be below some
scale $\Lambda$.  In this sense, $\Lambda$ acts as a radius of
convergence for the perturbative organization of the series.

\item {\em Each term in the Lagrangian has an unknown coefficient,
expected to be of order unity, which must be fit to data.}  Once the
effective Lagrangian has been truncated by the process described
above, a (hopefully small) number of such coefficients remain.  Of
course, the usefulness of the theory depends on few enough
coefficients remaining that the Lagrangian may then be used to predict
other observables.  The expectation that the coefficients are of order
unity once the known physics is taken into account is called the {\em
naturalness assumption}; if a coefficient turns out to be too small,
one suspects a hidden symmetry, while if it is too large, one suspects
that important physics has not been taken properly into account.

\end{enumerate}

	The relevance of this construction in the current context is
that a great deal of effort has recently been invested in developing
effective chiral theories to compute nucleon properties.  A nice talk
on the importance of chiral symmetry in nuclear interactions appears
in~\cite{Fri}, while~\cite{BKM} provides a very thorough review
through 1995.  The subtlety in the nucleon case is that the
development of the effective theory runs into complications because of
the presence of several mass scales.  For suppose, in the construction
described above, one finds not one but two scales of physics,
$\Lambda_1 \ll \Lambda_2$, relevant to a given process.  Then it is
not enough to merely choose processes with characteristic momenta $p$
satisfying $p \ll \Lambda_1$ and $p \ll \Lambda_2$, for the
combination $\Lambda_2 / \Lambda_1 \gg 1$ might appear in the
dimensionless unknown coefficients, making them unnaturally large and
thus defeating the predictivity of the theory.

	In the single nucleon case, in addition to the scale of the
onset of non-pionic interactions $\Lambda$, one must also deal with
the appearance of the nucleon mass, as discussed first in~\cite{Wein}.
One particularly successful treatment~\cite{JM2} is to use a
Foldy-Wouthuysen transformation\footnote{Actually, this is also how
the Heavy Quark Effective Theory is developed.  See, {\it e.g.},
Ref.~\cite{Geo}.} to remove nucleon mass terms from the Lagrangian, a
method that effectively replaces nuclear momenta with velocities.

	However, in the case of two or more nucleons, one typically
has a {\em three}-scale problem: momentum $p$, nucleon mass $M$, and
nuclear binding energy $p^2/2M$.  In this case, a typical approach is
to remove the scale $M$ as described above, and then to sum up
diagrams with the small scale $p^2/2M$ in nucleon propagators --- a
chain of loop diagrams --- using nonperturbative quantum mechanics in
the form of the Lippmann-Schwinger or Schr\"{o}dinger equation, into
an effective potential.~\cite{ORK} A very new approach~\cite{KSW}
eliminates the small binding scale by regularizing loop integrals
using minimal subtraction near $D=3$ dimensions rather than $D=4$, as
is usually done in field-theoretic calculations.  Then the loop
diagrams are summed by means of renormalization group equations, thus
avoiding the necessity of picking a kernel for a particular wave
equation.  That such an approach might work is perhaps not so
surprising: Binding energy scales are very small compared to the
nucleon masses, so the fundamental dynamics of the problem involves
perturbations about an essentially static nucleon, and therefore is
three-dimensional.

	Before leaving this topic, it should be pointed out that many
theories and models can be promoted to an effective theory.  All that
is needed is a set of symmetry principles for deciding what
interactions are allowed, and an organizing principle ({\it e.g.}, a
perturbation series) for deciding which of these interactions are
important.  For example, meson potential models have neglected
corrections in the form of nontrivial form factors or meson-meson
couplings, while quark potential models are typically organized in a
series in $1/m_{\rm quark}$.  All in all, the concept of the effective
theory is not unlike the famous Wigner-Eckart theorem.  Both divide
physics into a symmetry part and a dynamical part.  In the case of the
W-E theorem, the symmetry part is represented by spin SU(2)
Clebsch-Gordan coefficients, while the dynamical part is the so-called
reduced matrix element.  In effective theories, the symmetry part
consists of Lorentz, chiral, and parity invariances, and other
conditions we impose upon the interactions, while the dynamical part
is represented by the unknown coefficients that must be fit to data.
In this sense, effective theories are very minimal in their dynamical
content, but provide a very useful starting point for deeper inquiries
into the dynamics.

\subsection{Large $N_c$ QCD}

	It is a remarkable fact that considering the limit in which
the number $N_c$ of QCD color charges, which is 3 in our universe,
becomes infinite,~\cite{tH} actually simplifies strong interaction
physics.  How can {\em increasing\/} the number of degrees of freedom
actually lead to a simplification?  Think of statistical mechanics as
an analogy, where Avogadro's number of particles can be described by
just a few collective quantities, such as pressure, temperature, {\it
etc.}  In large $N_c$ QCD, baryons are treated similarly, in a
Hartree-Fock picture:~\cite{Witt} To first approximation, each of the
$N_c$ quarks feels only the collective effect of the other $N_c-1$.

	However, taking the large $N_c$ limit seriously means that one
expands physical quantities in a series in $1/N_c$.  If we apply this
to our universe, the expansion parameter is 1/3, which certainly does
not seem small!  However, for many quantities, the first correction to
the large $N_c$ limiting value appears not at relative order $1/N_c$
but $1/N_c^2 = 1/9$, which is arguably a small parameter.  Even if
this does not occur, one may simply adopt the expansion anyway, fit to
the data using the $1/N_c$ expansion and set $N_c = 3$ at the end of
the calculation.  Then one can see {\it a posteriori\/} whether the
factors of 1/3 truly are supported by experiment.  A simple example
was first pointed out in~\cite{Jenk} where it is observed that the
relative mass splitting between nucleons and $\Delta$ resonances is
suppressed by $1/N_c^2$.  Writing this relation in a scale-independent
way,
\begin{equation}
\frac{m_\Delta - m_N}{\frac 1 2 (m_\Delta + m_N)} =
O \left( \frac{J^2}{N_c^2} \right) .
\end{equation}
Experimentally, the l.h.s.\@ is 0.27, whereas the r.h.s.\@ is
3/$N_c^2$, which is 3 if we dismiss the factors of $N_c$ as
irrelevant, but $0.33$ if they are retained.  In fact, one can study
the entire spectrum of the ground state baryons this way,~\cite{JL}
and indeed the explicit factors of $N_c$ are essential to account
properly for all masses.

	In fact, studies of the large $N_c$ expansion for nuclear (as
opposed to nucleon) systems are in their infancy; only a handful of
papers studying this problem have yet appeared, but the prospects look
quite promising.  The basic lesson is that large $N_c$ provides a kind
of effective theory for nuclear systems, in that the old spin-flavor
SU(6) is known~\cite{DM} to hold in the large $N_c$ limit (the
symmetry), while interaction operators suppressed in this limit are
accompanied by powers of $1/N_c$ (the organizing principle).

	One direction that such a theory may be used is to note that,
if the leading interactions in $1/N_c$ obey some symmetry, then so do
the corresponding physical observables.  For example, nature obeys an
approximate symmetry under interchanges of the states $(p\uparrow ,
p\downarrow, n\uparrow, n\downarrow)$, the famous Wigner
supermultiplet.  In fact, this phenomenon has a large $N_c$
explanation~\cite{KS} in that the operators that would lift this
degeneracy are suppressed by powers of $1/N_c$.

	Another direction is to find exactly which operators appear at
leading order in $1/N_c$ for a given process, and study their symmetry
properties.  This is what is done in the first large $N_c$ analysis of
the NN interaction,~\cite{KM} where it is shown that one of the
leading operators acting on nucleons 1 and 2 is the combined
spin-isospin operator
\begin{equation} \label{tens}
\left( \mbox{\boldmath{$\sigma$}}_1 \cdot \mbox{\boldmath{$\sigma$}}_2
\right) \left( \mbox{\boldmath{$\tau$}}_1 \cdot
\mbox{\boldmath{$\tau$}}_2 \right) .
\end{equation}
It is important to realize that this sort of analysis is independent
of the particular dynamical origin of the given operator.  If one
assumes that the dynamics arises from one-meson exchanges, for
example, then one concludes that mesons with strong couplings to the
given operator will be important for the NN interaction.  In the given
example, we know that the $\pi$ and $\rho$ tensor couplings contain
pieces like (\ref{tens}), and so large $N_c$ explains why we might
have expected that meson potential models require large tensor
couplings to these mesons.  But any model, meson exchange or not, that
successfully describes the NN interaction must recognize the
importance of nucleon operators such as (\ref{tens}).

\section{Conclusions}

	In the final analysis, we return to a variant of our original
question: Why does the meson exchange picture work so well for the NN
interaction, when the underlying theory of QCD is so much more
complicated?  The ultimate contribution of QCD to the understanding of
the NN interaction will almost certainly not be in the form of a
solution to some as yet unknown field equation, but rather the
realization of how a complicated collection of quarks and gluons
possesses some limiting case in which the system achieves a collective
degree of simplicity, which we observe as a pair of interacting
nucleons.  We have already begun to see such simplifications take
place in effective theories, and especially in large $N_c$ QCD.

	Even though we cannot yet solve the strong interaction
problem, we have begun to nibble at the edges.  For example, we have
argued that the famous ``repulsive core'' of the NN potential appears
to be due to quark effects.  It would be exciting to find more
evidence for a six-quark ``bag'' or colored particle exchanges as QCD
suggests, phenomena that are quite exotic from the one-meson exchange
perspective.  Obviously these are topics of interest to both the
nuclear and particle communities.

	This last observation lies at the crux of my optimism on the
future of NN studies: After following divergent paths for some
decades, nuclear and particle physics are again making great strides
together.  We will see much more of the fruits of this combined effort
in the future.

\section*{Acknowledgments}
I would like to extend a special {\it d\u{e}kuji v\'{a}m\/} to the
conference organizers for their kind invitation and hospitality, and
to Franz Gross and Wally van Orden for valuable comments on the
content of the talk.  This work was supported by the U.S. Department
of Energy under contract No.\ DE-AC05-84ER40150.

\end{document}